\documentclass[useAMS,usegraphicx]{mn2e}
\usepackage{rotate}
\usepackage{times}
\newif\ifAMStwofonts
\AMStwofontstrue

\def\gsim{\mathrel{\hbox{\rlap{\hbox{\lower4pt\hbox{$\sim$}}}\hbox{$>$}}}}
\def\lsim{\mathrel{\hbox{\rlap{\hbox{\lower4pt\hbox{$\sim$}}}\hbox{$<$}}}}

\def\xmm{{\it XMM-Newton}}
\def\chandra{{\it Chandra}}

\def\asca{{\it ASCA}}
\def\sax{{\it BeppoSAX}}

\def\xmm{{\it XMM-Newton}}

\def\et{{et al.\ }}
\def\mcg{{MCG--6-30-15}}

\def\cm{{\rm\thinspace cm}}

\title{A long hard look at \mcg\ with \xmm}

\author[Fabian \et]
       {A. C. Fabian,$^{1}$\thanks{E-mail: acf@ast.cam.ac.uk},
        S. Vaughan,$^{1}$
        K. Nandra,$^{2,3}$
        K. Iwasawa,$^{1}$
        D. R. Ballantyne,$^{1}$
        J. C. Lee,$^{4}$
\newauthor   A. De Rosa,$^{1, 5}$
        A. Turner$^{1}$ and
        A. J. Young$^{6}$ \\
$^{1}$Institute of Astronomy, University of Cambridge, Madingley Road, Cambridge CB3 0HA\\
$^{2}$Laboratory for High Energy Astrophysics,  NASA/Goddard Space Flight Center, Greenbelt, MD 20771, USA\\
$^{3}$Universities Space Research Association\\
$^{4}$Massachusetts Institute of Technology, Center for Space Research, 77 Massachusetts Ave. NE80, Cambridge, MA 02139, USA\\
$^{5}$IASF/CNR, Roma, Italy\\
$^{6}$Astronomy Department, University of Maryland, College Park, MD 20742, USA\\
}
\date{Accepted 6/6/2002; submitted 30/5/2002; in original form 3/4/2002}
\pagerange{\pageref{firstpage}--\pageref{lastpage}}
\pubyear{2002}
\begin{document}
\maketitle
\label{firstpage}

\begin{abstract}
We present first results from a 325~ks observation of the
Seyfert 1 galaxy \mcg\ with \xmm\ and \sax. The
strong, broad, skewed iron line is clearly detected and is well
characterised by a steep emissivity profile within $6r_{\rm g}$ (i.e.
$6GM/c^2$) and a flatter profile beyond. The inner radius of the
emission appears to lie at about $2r_{\rm g}$, consistent with results
reported from both an earlier \xmm\ observation of \mcg\
by Wilms \et and part of an \asca\ observation by Iwasawa \et when
the source was in a lower flux state. The radius and steep emissivity
profile do depend however on an assumed incident power-law continuum
and a lack of complex absorption above 2.5~keV. The blue wing of the
line profile is indented, either by absorption at about 6.7~keV or by
a hydrogenic iron emission line. The broad iron line flux does not follow
the continuum variations in a simple manner. 
\end{abstract}


\section{Introduction}

The Seyfert 1 galaxy \mcg\ has played an important role in studies of
accretion onto black holes due to the presence of a broad, skewed iron
line in its X-ray spectrum (Tanaka \et 1995). The shape of the line
seen with \asca\ is consistent with emission from the surface of an
accretion disc extending from about 6 to more than 40 gravitational
radii ($6-40r_{\rm g}; r_{\rm g}=GM/c^2$) inclined at about 30 deg to the
line of sight (Fabian \et 1995). Occasionally the red (i.e. lower
energy) wing of the line is seen to extend below 4~keV (Iwasawa \et
1996; 1999). This can be explained by the disc extending within
$6r_{\rm g}$ which may imply the black hole must be rapidly
spinning. The presence of the broad iron line in \mcg\ has been
confirmed by \sax\ (Guainazzi \et 1999), \xmm\ (Wilms \et 2001) and
\chandra\ (Lee \et 2002).

Here we present preliminary results from a long 325~ks observation of
\mcg\ made with \xmm. The source was at a similar flux level to the
previous \asca\ observations, and about 70 per cent brighter than during the
earlier 100~ks \xmm\ observation reported by Branduardi-Raymont \et
(2001) and Wilms \et (2001). Simultaneous observations were made with
\sax, providing coverage from $\sim$0.2--100~keV. 
The present work focuses on the spectrum above 2.5~keV and the iron-K line
features; absorption and emission features below 2~keV due to oxygen
and other elements will be discussed more fully in later work.

\section{Data Reduction}

\mcg\ was observed by \xmm\ (Jansen \et 2001) over the period 2001
July 31 -- 2001 August 5 (rev. 301, 302 and 303), during which
all instruments were operating nominally. The present analysis is
restricted to the data from the European Photon Imaging Cameras (EPIC).
Both the EPIC MOS cameras (Turner \et 2001) and the EPIC pn camera
(Str\"{u}der \et 2001) were operated in small window mode and 
used the medium filter. Extraction of science products from the
Observation Data Files (ODFs) followed standard procedures using the
\xmm\ Science Analysis System v5.2 (SAS).

The EPIC data were processed using the standard SAS processing chains.
Source data were extracted from circular regions of radius 30 arcsec
from the processed MOS and pn images and background events were
extracted from regions in the small window least effected by source
photons.  These showed the background to be relatively low and stable
throughout the observation, with the exception of the final few ks of
each revolution where the background rate increased. Data from these
periods were ignored.   The total amount of ``good'' exposure time
selected was 315~ks and 227~ks for MOS and pn, respectively. (The
lower pn exposure is due to the lower ``live time'' of the pn camera
in small-window mode, $\sim$71 per cent).

The ratios of event patterns as a function of energy  showed there is
negligible pile-up in the pn data but the MOS data suffer slightly
from pile-up.   A comparison of the data extracted including and
excluding events from the central 10 arcsec  of the source region
showed only a slight flattening ($\Delta \Gamma \sim 0.06$) in the
piled-up spectrum.  This small effect was ignored in order to
increase the number of source counts.

Events corresponding to patterns 0--12 (single--quadruple pixel) were
extracted from the MOS cameras and patterns 0--4 (single+double) were
used for the pn spectral analysis, after checking for consistency with
the pattern 0 (single pixel event) data.  The total number of photons
extracted from the source region was $6.23 \times 10^{6}$ for the pn
and $2.35 \times 10^{6}$ for each MOS camera.  The \sax\ data were
processed in the standard manner (Fiore \et 1999).

\begin{figure}
\rotatebox{270}{
\resizebox{!}{\columnwidth}
{\includegraphics{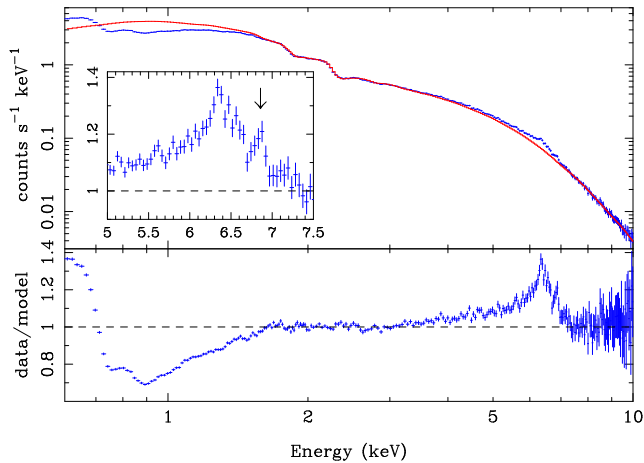}}}
\caption{
Combined MOS spectrum (the data were combined for plotting purposes
only) and ratio of data to a power-law model joining the 2--3 keV data
and 7.5--10 keV data. As this is not a realistic model for the
continuum the residuals should be considered merely as representative
of the spectral complexity.  The inset panel shows a close-up of the
iron line region with the 6.9 keV feature marked.
}
\label{fig:mos_spectrum}
\end{figure}

\section{Spectral Analysis}
\label{sect:spec}

The spectral calibration of the EPIC cameras is still evolving. Using
the publicly available software and response matrices,  the remaining
calibration-dependent residuals over the 2.5--10~keV range considered
here should be at the $\lsim 5$ per cent level (see Ferrando \et 2002).
However, there
remain problems with the model for charge transfer inefficiency (CTI)
for pn small-window mode data. In order to mitigate
the effects of this, a conservative approach was used for the present
analysis, with detailed spectral fitting results based on only the MOS
data.

The source spectra were grouped such that each bin contains
at least 20 counts and were fitted using {\sc XSPEC v11.1} 
(Arnaud 1996).  The spectral responses used 
were {\tt m1\_medv9q19t5r5\_all\_15.rsp} for MOS1 (and
similarly for MOS2) and  {\tt epn\_sw20\_sdY9\_medium.rmf} for the pn.
The quoted errors on the derived model parameters
correspond to a 90 per cent confidence level for one interesting
parameter (i.e. a $\Delta \chi^{2}=2.7$
criterion), unless otherwise stated,
and fit parameters (specifically line and edge energies) are quoted
for  the rest frame of the source. 

The EPIC MOS data show a spectral form remarkably similar to those
seen in earlier \asca\ observations, as illustrated in
Fig~\ref{fig:mos_spectrum} (cf. Fig.~1 of Iwasawa \et 1996).  The
residuals around 6~keV have been interpreted as a broadened and
redshifted iron K$\alpha$ emission from a relativistic accretion disc
(Tanaka \et 1995).  Only the \xmm\ EPIC
spectrum above 2.5~keV is examined here, to limit the effects of the complex soft
X-ray absorption (see also Lee \et 2001), a detailed treatment of
which is beyond the scope of this Letter. This energy range also
excluded the region around the detector Si K-edge (1.84~keV) and the
mirror Au M-edge (2.3~keV).  To better constrain the continuum, and in
particular the Compton reflection component (e.g. George \& Fabian
1991; Nandra \& Pounds 1994), the \sax\
MECS and PDS data were also fitted simultaneously. 
The normalisations of the MOS and MECS data were
allowed to vary independently, to account for differences in the
absolute calibration, with the PDS normalisation tied to 0.86
times the MECS value (Fiore \et 1999). Galactic absorption 
($N_{\rm H}=4.06 \times 10^{20}$~cm$^{-2}$; Dickey \& Lockman 1990) was
included  in all the models discussed below.

\begin{table*}
\centering
\caption{Results of simultaneous fits to the \xmm\ MOS and \sax\
data. Line energies ($E$) and widths ($\sigma$) are given in units of keV and
equivalent widths ($EW$) are given in eV. Disc radii (e.g. $R_{\rm
in}$) are given in units of $r_{\rm g}$ and inclination ($i$) is
given in degrees. The models are described in section~\ref{sect:spec}. \label{tab:fits}}
\begin{center}
\begin{tabular}{lccccccccc}                
\hline
model & \multicolumn{1}{c}{Continuum} & \multicolumn{3}{c}{Broad K$\alpha$} &\multicolumn{3}{c}{Core K$\alpha$}  &  &$\chi^{2}/dof$ \\
 & $\Gamma$ & $E$ & $\sigma$ & $EW$ &$E$ & $\sigma$ & $EW$ & &  \\
1 &$1.93\pm0.01$ &$4.29_{-0.28}^{+0.09}$ &$1.72\pm0.09$ &$383_{-70}^{+25}$ &$6.40\pm0.03$ &$0.27\pm0.06$ &$119_{-13}^{+9}$ & & 888.2/1033  \\
\hline
model & \multicolumn{1}{c}{Continuum} & \multicolumn{4}{c}{Broad K$\alpha$} & & & &$\chi^{2}/dof$  \\
 & $\Gamma$ & $i$ & $R_{\rm in}$ & $q$ & $EW$ &  &  & &  \\
2 &$1.92\pm0.01$ & $38.3\pm1.2$ &$4.6\pm0.4$ & $5.6_{-0.6}^{+0.3}$&$308\pm25$ & & & & 979.4/1034\\
\hline
model & \multicolumn{1}{c}{Continuum} & \multicolumn{6}{c}{Broad K$\alpha$} & 6.9 keV line &$\chi^{2}/dof$ \\
 & $\Gamma$ & $i$ & $R_{\rm in}$ & $R_{\rm br}$ &$q_{\rm in}$  &$q_{\rm out}$  & $EW$ & $EW$ & \\
3 &$1.96\pm0.01$ &$28.4\pm1.0$ & $2.0\pm0.1$&$6.2_{-1.4}^{+3.4}$ &$4.9\pm0.6$ &$2.6\pm0.3$ &$554\pm56$ &$29\pm5$ & 885.1/1031\\
\hline
4$^{1}$ & $1.95\pm0.04$ & $27.8\pm1.3$ & $2.0\pm0.2$ &
 $6.5_{-1.4}^{+4.5}$ & $4.8\pm0.7$ & $2.5_{-0.4}^{+0.2}$ & $413_{-89}^{+73}$ & $21\pm5$ & 856.2/1029 \\
\hline
\end{tabular}
\end{center}

\raggedright
$^{1}$ This model includes  $R_{\rm refl}$ and $E_{\rm fold}$ as free parameters,
$A_{\rm Fe}=3$ and emission from Fe K$\beta$ (see section~\ref{sect:complex}).
\end{table*}

\subsection{Phenomenological model}

The iron line residuals were initially parameterised using a simple model
comprising a power law plus Compton reflection, narrow
($\sigma=10$~eV) Gaussian at 6.4~keV and a broad Gaussian whose energy
and width were free parameters. The
Compton reflector was assumed to be a neutral, static slab subtending
a solid angle of $2\pi$~sr at the X-ray source, inclined at
$i=30$~deg, and having solar
elemental abundances (Magdziarz \& Zdziarski 1995). The primary
continuum was a power law with an exponential cutoff at a fixed energy
of 400~keV (constraints on the reflection and
continuum parameters are discussed later). This model gave a good overall fit with
$\chi^{2}=984.1$ for 1034 degrees of freedom ($dof$).  The broad line
component had an energy $E=4.8\pm0.3$~keV, a width of $\sigma=1.8\pm
0.2$~keV and equivalent width $EW=527\pm 70$~eV.  The narrow, 6.4~keV
``core'' had $EW=38\pm10$~eV and peaked very close to 6.4~keV, with
$E=6.40\pm 0.04$~keV if allowed to be free.

Very narrow components to the iron line, unresolvable by EPIC, have
been observed in the \chandra\ HETGS spectra of several Seyferts
(Yaqoob \et 2002). They may be identified with emission from,
e.g. the BLR or molecular torus. Such narrow lines are not ubiquitous,
however, and allowing the 6.4~keV component to have a finite
width is resolved with a width corresponding to $FWHM\approx
30,000$~km s$^{-1}$ and $EW=120$~eV. The fit is excellent, the
parameters are given in Table~\ref{tab:fits} (model
1). A preliminary analysis of the pn data gave generally
consistent results and in particular the core of the line was
again resolved. See also Lee \et (2002).
The large width of the line core suggests that this also
arises in the disc, rather than at a greater distance (cf. NGC
5548; Yaqoob \et 2001). 

\subsection{Relativistic line profiles}

To provide a more physically realistic description of the data 
the line profile of Laor (1991) was used to describe 
a broad line from an accretion disc
surrounding a rotating black hole. 
The disc was assumed to be in a
low state of ionisation, with the rest energy of the line fixed at
6.4~keV. The fits were relatively insensitive to the outer radius of
the disc, which was therefore fixed at $R_{\rm out}=400 r_{\rm g}$.  The free
parameters in the model were: the inner radius of the disc, $R_{\rm
in}$, the inclination, $i$, and the emissivity as a function of
radius, which was parameterised as a power law of index $q$ (i.e.
$R^{-q}$). 
Both the line and the reflection  spectrum were blurred
using the kernel from the {\sc laor} model. The fit parameters are
given in Table~\ref{tab:fits} (model 2), the $\chi^{2}$ is comparable
to the broad plus narrow Gaussian model, but falls short of the best
fitting $\chi^{2}$ with two broad Gaussians (model 1). The reason for
this is that, with a simple power law emissivity function, the Kerr
line profile cannot simultaneously model the broad, red wing of the
line and the narrower core around 6.4~keV. The preliminary pn
analysis gave comparable results but suggested an even stronger red
wing to the line.

\subsection{Additional iron K components}

Before exploring the red wing further it should be noted that there is some
additional complexity in the line region at $\sim 7$~keV (see inset of
Fig~\ref{fig:mos_spectrum}). Including an iron K-edge in model 2
provided a significant improvement in the fit ($\chi^{2}=908.2/1032$),
with an edge energy of $E=7.38\pm0.09$~keV and a depth of
$\tau=0.11\pm0.02$, similar to that discussed by Pounds \& Reeves
(2002).  However, such an edge must have a physical origin; if it is
due to neutral or near-neutral iron along the line of sight, then
strong iron-L absorption is expected which is inconsistent with the
soft X-ray spectrum (Lee \et 2001). If the material is out of the line
of sight and the edge is due to reflection off distant material, the
reflection component must be very strong ($R\sim 3$), which is then
inconsistent with the relatively weak, narrow 6.4~keV iron line
observed in the spectrum.  Another possibility is that the edge arises
from thick material partially covering the line of sight. This gives a
comparable fit ($\chi^{2}=894.6/1032$) to the edge, with a covering
fraction of $\sim 0.16$. Using this model the  broad line parameters
differed only slightly from those of model~2, and the broad, red
wing of the line, extending within $6r_{\rm g}$, was still required in
the fit. 

Sako \et (2002) suggested there may be significant  absorption in the
6.4--6.7~keV range due to inner-shell transitions of ionised
iron. This possibility was explored by including an absorption line in
model 2. This gave a good fit ($\chi^{2}=923.9/1031~dof$) with
absorption line parameters $E=6.74\pm0.05$~keV and
$EW=138\pm35$~eV. The relativistic line parameters differ only
slightly, with the emissivity index changing to $q=6.6\pm1.1$.

An alternative interpretation is that the complexity around 7.0~keV is
in an emission component, identified with recombination emission by
H-like iron. Adding a Gaussian to model 2 at $E\approx 6.9$~keV
again improved the fit ($\chi^{2}=956.1/1032~dof$) and
gives best fitting parameters of $E=6.91\pm0.03$~keV and
$EW=18\pm6$~eV.  If a line of this strength is due to gas photoionised
by the observed continuum then it must lie within a radius of $\sim
2\times 10^{16}\cm$. Since it is not possible to unambiguously 
determine between absorption and emission
as the source of complexity above 6.5~keV, both are considered
in the following subsections.

\subsection{More complex models}
\label{sect:complex}

A steep emissivity function was required to model the red wing of the line (see
also Wilms \et 2001), but this gave too little flux from the outer
parts of the disc to model the core of the line.  This raises the
possibility that the emissivity function is not a simple power-law.
This was explored by using a broken power-law for the
emissivity function, with emissivity index $q_{\rm in}$ from $R_{\rm in}$ to
$R_{\rm br}$ and index $q_{\rm out}$ from $R_{\rm br}$ to $R_{\rm out}$. The
fit parameters for this model are given in Table~\ref{tab:fits} (model
3).

\begin{figure}
\rotatebox{270}{
\resizebox{!}{\columnwidth}
{\includegraphics{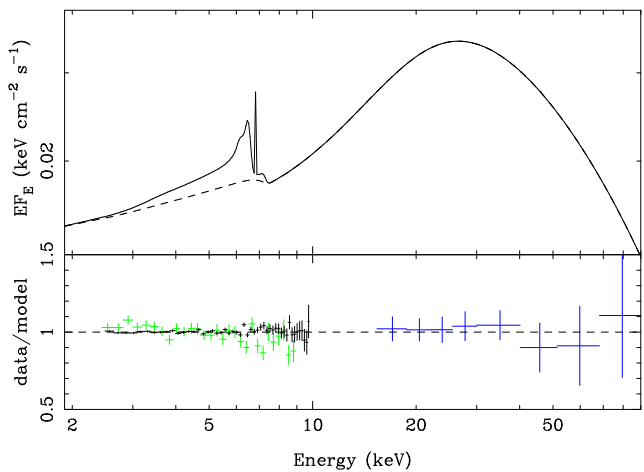}}}
\caption{Model including relativistic iron line emission (model 4)
and fit residuals for \xmm\ MOS (black) and \sax\ MECS (green) and PDS
(blue).}
\label{fig:spectrum_with_sax}
\end{figure}

This fit is not physically self-consistent for at least two reasons.
Firstly it ignores the emission from Fe K$\beta$, which is expected at
7.05~keV. Secondly, the high $EW$ of the line is not consistent
with our assumptions about the reflection component, namely
that it has $R_{\rm refl}=1$, $i=30$~deg, solar abundances and is
neutral. In such circumstances, considering the observed value of
$\Gamma$, the predicted $EW$ is only 140~eV (George \&
Fabian 1991). Two plausible possibilities are that the iron abundance
in the accretion disc is higher than the solar value (see e.g. Lee \et
1999), or that the reflection component is enhanced for some
reason, for example due to anisotropy (Ghisellini \et 1991) or
geometry (Fabian \et 2002). 

\begin{figure}
\rotatebox{270}{
\resizebox{!}{\columnwidth}
{\includegraphics{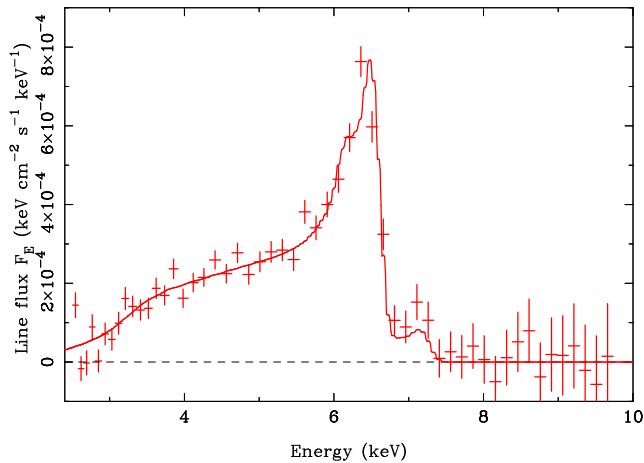}}}
\caption{
Relativistic iron line profile,
shown in ($F_{\nu}$) flux units, obtained from the ratio of the MOS data
to best-fitting underlying continuum model (model 4), multiplied by the
continuum model in flux units (as opposed to an ``unfolded'' plot).
The crosses mark the data points and 
the solid line marks the line model. }
\label{fig:fluxed_line}
\end{figure}

The high abundance case was examined by fixing the iron abundance of
the reflector to $3\times$solar in model 3.  This in fact improved the
overall fit compared to the solar abundance case ($\chi^2 =
877.2/1031~dof$). Changes in the disc line parameters were negligible.
The value of the cutoff energy can affect both the high energy
spectral shape and the apparent strength of the reflection component.
Allowing $E_{\rm cut}$ to be free in this fit gave $E_{\rm
cut}=241_{-86}^{+228}$~keV ($\chi^2=875.4/1030~dof$).   
Allowing $R_{\rm refl}$ to be a free parameter further improved the
fit, therefore this was also left free in model 4.

Model 4 comprised a power-law plus cold reflection (with
overabundant iron, $A_{\rm Fe}=3$),
emission from iron K$\alpha$ (6.40~keV) and K$\beta$ (7.05~keV). The
reflection plus line spectrum was then blurred using the {\sc laor}
kernel (with a broken power-law emissivity function). A narrow
emission line at $E=6.9$~keV (fixed) was also included in the model.
This provided the best fitting model, with a cut-off energy
$E_{\rm cut}=122_{-18}^{+90}$~keV and $R_{\rm
refl}=2.2_{-0.7}^{+1.1}$, the other  parameters are given in 
Table~\ref{tab:fits}. The model and fit residuals are
shown in Fig.~\ref{fig:spectrum_with_sax} and the ``fluxed'' line
profile based on this fit is shown in Fig.~\ref{fig:fluxed_line}.

Including an absorption line instead of a 6.9~keV emission line gave a
slightly poorer fit ($\chi^{2}=863.0/1027~dof$) with line parameters
$E=6.74\pm0.02$~keV,  $EW=31\pm9$~eV (consistent with the prediction
of Sako \et 2002), and slight changes to the
relativistic line parameters (in particular $R_{\rm br}=3.9\pm0.7r_{\rm g}$).

\section{Spectral Variability}
\label{sect:spec_var}

\begin{figure}
\rotatebox{270}{
\resizebox{!}{\columnwidth}
{\includegraphics{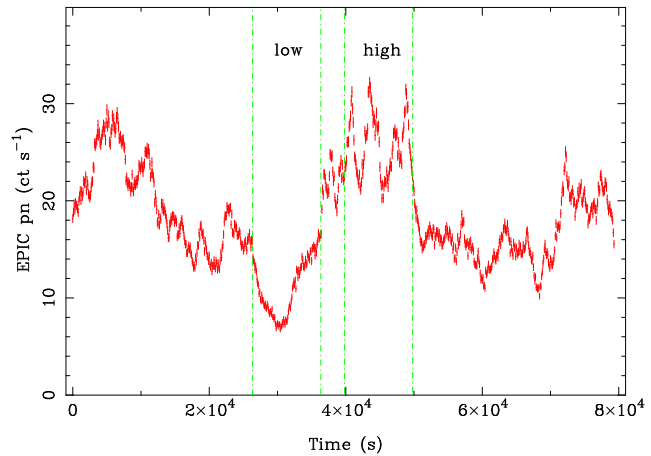}}}
\caption{
EPIC pn light curve (0.2--10~keV) for revolution 301 showing the
intervals of low and high flux discussed in section~\ref{sect:spec_var}. 
}
\label{fig:one_orbit}
\end{figure}

\begin{figure}
\rotatebox{270}{
\resizebox{!}{\columnwidth}
{\includegraphics{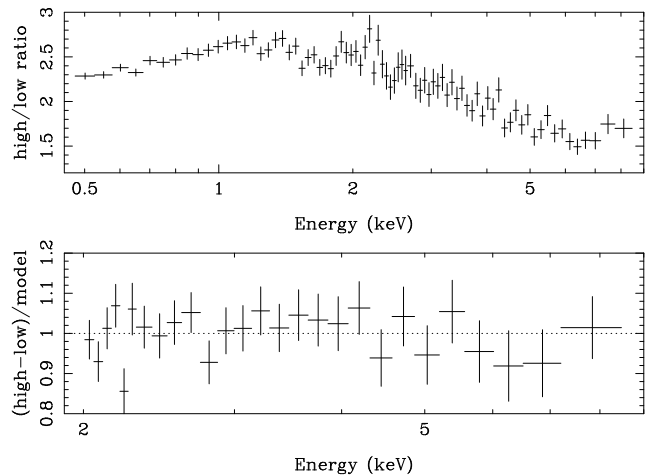}}}
\caption{Comparison of the high-flux and low-flux pn spectra extracted
from intervals marked on Fig.~\ref{fig:one_orbit}. Top panel: ratio
of high-flux to low-flux spectrum. 
Bottom panel: comparison of the difference spectrum (high--low) to a
power-law model.}
\label{fig:difference}
\end{figure}

As a first test for spectral variability around the iron line region
pn spectra were extracted from 10~ks intervals around the
minimum and maximum flux periods during revolution 301 (see
Fig.~\ref{fig:one_orbit}). The average continuum fluxes during these
intervals differ by $>\times 2$. The ratio of the two spectra is shown
in the top panel of Fig.~\ref{fig:difference} and shows clear signs
of spectral variability. There is a slight depression around 6.4~keV,
indicating the iron line core is stronger (compared to the continuum) in
the low-flux spectrum. The bottom panel of Fig.~\ref{fig:difference}
shows the residuals from fitting a power-law model to the difference
spectrum (i.e. high-flux minus low-flux spectrum). The difference
spectrum is consistent with a power-law in the range 2-10~keV and
indicates the iron line flux changed little between the two intervals.
A detailed analysis of the spectral variability properties 
will be given in a forthcoming paper but these first results are
consistent with the simple two component continuum model described in
Shih \et (2002). 

\begin{figure}
\rotatebox{270}{
\resizebox{!}{\columnwidth}
{\includegraphics{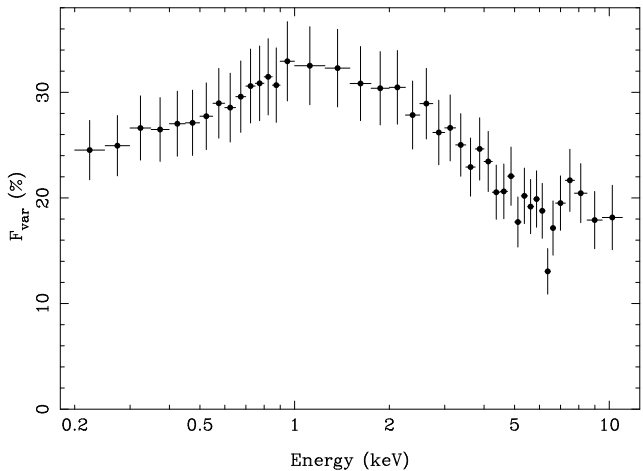}}}
\caption{The RMS spectrum based on rev. 302 EPIC pn data. The errors are
calculated as in Edelson \et (2002) and should be considered
conservative estimates of the true uncertainty.}
\label{fig:rms}
\end{figure}

The suggestion that the iron line was not as variable as the continuum
was borne out by an examination of the RMS
spectrum. Figure~\ref{fig:rms}  shows the normalised RMS variability
spectrum derived from the revolution 302 pn data.  The RMS spectrum
clearly shows that the fractional variability amplitude is suppressed
at energies close to the iron line, compared to surrounding continuum
bands (see also Inoue \& Matsumoto 2001).

A further point to note is that the RMS spectrum (and the ratio of
high/low spectra) are smooth in the 0.2--1.0~keV range. In particular,
the absence of any feature at 0.7~keV in the spectral ratio is
expected if the large drop at that energy (Fig~\ref{fig:mos_spectrum}) 
is due to absorption (with constant
optical depth). For the drop to be the blue wing of an emission line 
(Branduardi-Raymont \et 2001),
the line intensity must be responding linearly to the continuum flux
(unlike the iron line). An examination of the RGS spectra confirms this
result.

\section{Discussion}
\label{sect:disco}

A long observation with \xmm\ and \sax\ of \mcg\ in its typical state
has again confirmed the presence of the broad, skewed iron line.  All
the models considered above are formally acceptable
($\chi_{\nu}^{2}<1.0$), and all include emission within  $6r_{\rm g}$,
consistent with emission from a disc around a spinning black
hole.

Model 4 provides the best (in a $\chi^2$-sense) and most physically
self-consistent explanation of the data.  In this model, the disc
emissivity is described by a broken power-law in radius, and it is
interesting to note that the break radius occurs at $\sim 6r_{\rm
g}$. Beyond  this radius the disc has an  emissivity profile $q_{\rm
out} \sim 2.5$ and produces an iron line with an equivalent width
$\sim 200$~eV (the $5.5-6.5$~keV core of the line shown in
Fig.~\ref{fig:fluxed_line}), both close to the values expected from
standard accretion disc models. Within $6r_{\rm g}$ the emissivity
steepens, producing the strong low-energy tail to the line emission,
also with an equivalent width $\sim 200$~eV, suggesting additional
physics within this region beyond that expected from standard
accretion disc models (see Wilms \et 2001). The strong iron line
is consistent with an enhanced reflection spectrum and an
overabundance of iron or ionisation of the disc surface.
This last possibility will be examined in a later paper.

The iron line parameters do, however, depend slightly on the assumptions made
about the complex absorption.  In particular all the above models
formally assume the effects of ionised absorption are negligible above
2.5~keV, which is correct for absorption due to low-Z elements (e.g.
O) but may not be true if there is absorption by higher-Z elements
(e.g. Si, S). A preliminary analysis allowing for the possibility of
Si and S edges did not change the requirement for a strong red wing to
the iron line. A detailed analysis of the RGS data will yield
constraints on the warm absorption (particularly on the low Z
elements) and help remove existing degeneracies. 

Future work to determine the spin must include emission from the
immediate plunge region inside the innermost stable orbit (Reynolds \&
Begelman 1997; Agol \& Krolik 2000) as well as returning radiation
(Martocchia, Matt \& Karas 2002).

\section*{Acknowledgements}
Based on observations obtained with \xmm, an ESA science mission with
instruments and contributions directly funded by ESA Member States and
the USA (NASA). ACF thanks the Royal Society for support.
\label{lastpage}

\end{document}